\documentclass[twocolumn,preprintnumbers,pra,superscriptaddress,showkeys]{revtex4}

\usepackage{amsmath,amssymb,amsfonts,graphicx,float,times,psfrag}
\usepackage[pdftex]{color}
\usepackage{amsmath,bm}
\usepackage[colorlinks, linkcolor=blue, citecolor=blue,  breaklinks=true]{hyperref}
\usepackage{mathtools,amsfonts,mathptmx}
\usepackage[utf8]{inputenc}
\usepackage[T1]{fontenc}
\usepackage{xcolor}
\usepackage{braket}
\usepackage[titletoc,title]{appendix}
\usepackage[nameinlink,capitalize]{cleveref}

\usepackage[shortlabels]{enumitem}
\begin{document}
\title{Quadratically enhancing optomechanical entanglement via dark mode control} 

\author{A.-H. Abdel-Aty}
\email{amabdelaty@ub.edu.sa}
\affiliation{Department of Physics, College of Sciences, University of Bisha, Bisha 61922, Saudi Arabia}

\author{D. R. K. Massembele}
\affiliation{Department of Physics, Faculty of Science, 
University of Ngaoundere, P.O. Box 454, Ngaoundere, Cameroon}

\author{P. Djorwé}
\email{djorwepp@gmail.com}
\affiliation{Department of Physics, Faculty of Science, 
University of Ngaoundere, P.O. Box 454, Ngaoundere, Cameroon}
\affiliation{Stellenbosch Institute for Advanced Study (STIAS), Wallenberg Research Centre at Stellenbosch University, Stellenbosch 7600, South Africa}

\author{A. N. Al-Ahmadi}
\affiliation{Department of Physics, Faculty of Sciences, Umm Al-Qura University, Makkah, 24382, Saudi Arabia}

\author{K.S. Nisar}
\affiliation{Department of Mathematics, College of Science and Humanities in Alkharj, Prince Sattam Bin Abdulaziz University, Alkharj 11942, Saudi Arabia}

\begin{abstract}
We propose a scheme to enhance quantum entanglement in an optomechanical system consisting of two mechanically coupled mechanical resonators, which are driven by a common electromagnetic field. Each mechanical resonator is linearly and quadratically coupled to the electromagnetic field. Moreover, the mechanical coupling between the resonators is modulated through a given phase that allows interference control in our structure. By tuning this phase, our system exhibits interference like-structure which is reminiscent of bright and dark mode features. The breaking of the dark mode via the phase adjustment leads to an entanglement generation, which is greatly enhanced through the quadratic coupling. Furthermore, the generated entanglement is robust enough against thermal noise and this resilience is improved when the quadratic coupling is accounted. Our work provides a way to enhance quantum entanglement via quadratic coupling which is assisted by interference control. Such quantum resources can be useful for quantum information processing, quantum computing, and other numerous quantum tasks. 
\end{abstract}

\maketitle
\date{\today}

\section{Introduction}\label{intro}
Quantum entanglement is an interesting resource that is useful for number of quantum technologies such as quantum information  processing \cite{Wendin.2017,Slussarenko.2019,Meher.2022}, quantum computing \cite{Zidan2021}, quantum sensing \cite{Degen_2017,Stray2022} and metrology \cite{Polino_2020,Barbieri2022}, just to name few.  However, engineering entangled states is not an easy task, and sometime these states are fragile to decoherence induced by thermal fluctuations owing to the interaction they have with their environment. While many platforms have been used to generate entangled states \cite{Thomas2020,Genes.Bri.2024}, optomechanical systems have been revealed as promissing candidates to synthetize robust and stable nonclassical states \cite{Wise_2024,Agasti2024,Djo2024}. Beside the generation of quantum correlations, optomechanical systems have been also used to improve plethora of technological applications such as quantum information \cite{Stannigel_2012}, sensing \cite{Qin_2019,Xia_2023,Mao_2023,Djorwe.2019,Tchounda_2023,Dj2024}, and those based on chaos \cite{Madiot_2021,Zhu_2023,Stella_2023} and synchronization \cite{Rodrigues_2021,Li_2022,Djorwe.2020,Sun_2024} among others.  

Several technics have been used in optomechanics to engineer entangled states. Among these schemes, there are those based on  Duffing nonlinearity \cite{Rostand2024}, Kerr nonlinearity \cite{Zhang_2017,Cha_2017}, disspative coupling \cite{Liao_2019,Zhang2021,Neto2022},  and quadratic coupling \cite{Garg_2023,Liu_2024}. Beside these configurations, it has been recently shown that multipartite optomechanical structures may host dark modes, which are detrimental for entanglement generation. In this case, breaking of that dark mode leads to entanglement enhancement in the system. Such schemes based on dark mode control in order to generate entanglement have been implemented in \cite{Lai_2022,Nori2022}, and strong entangled states were engineered. 

In this work, we aim to engineer entanglement in optomechanical system by proposing a scheme that exploits joint effect of both quadratic coupling and dark mode control. In our knowledge, such a proposal has not yet been investigated in the literature. Our benchmark system consists of two mechanically coupled mechanical resonators, which are driven by a common electromagnetic field. Moreover, the mechanical coupling that is captured by the phonon hopping rate $J_m$, is phase modulated as recently proposed \cite{Fang.2017,Mathew_2020,Wang_2020}. By tunning this phase, the dark mode can be controlled and broken. The breaking of this dark mode induces entanglement, and this induced entanglement is greatly enhanced thanks to the quadratic coupling in our proposal. Moreover, this generated entanglement exhibits oscillatory behavior over the modulated phase, and this behavior is reminiscent of entanglement death and revival which is an interesting prerequisite to synthetize entanglement. Furthermore, our generated entanglement is robust enough against thermal noise in the presence of quadratic coupling. Our proposal can be implemented in electromechanical system, in optomechanical structure, and in hybrid configuration. Our work provides an alternative way to enhance quantum entanglement by taking advantage of joint effect of both quadratic coupling and dark mode control.     

The rest of the manuscript is organized as follow. \autoref{sec:model} presents the model and derives the dynamical equations. Moreover, this section analyzes the stability of our system. The entanglement dynamics and its enhancement through the quadratic coupling assisted by dark mode control is presented in \autoref{sec:entang}. Our work is concluded in \autoref{sec:concl}.  

\section{Dynamical equations and system stability} \label{sec:model}

\begin{figure}[tbh]
  \begin{center}
  \resizebox{0.5\textwidth}{!}{
  \includegraphics{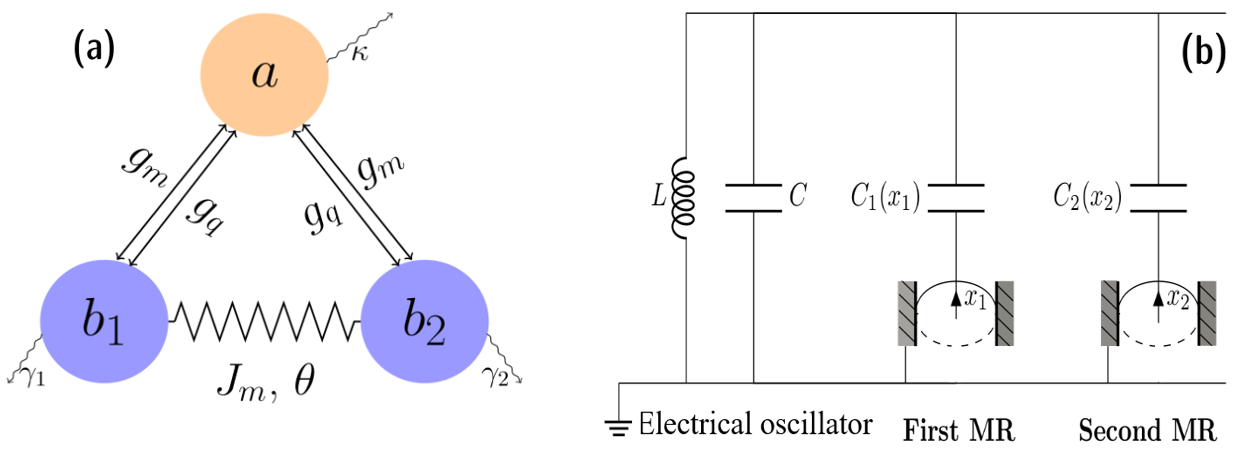}}
  \end{center}
\caption{Sketch of our benchmark system. (a) Optomechanical structure in which two mechanically coupled mechanical resonators are driven by a common optical field. Each mechanical resonator is both linearly and quadratically coupled to the driving field. (b) A possible equivalent electromechanical version of our optomechanical structure.}
\label{fig:Fig1}
\end{figure}

Our benchmark optomechanical system consists of two mechanical resonators driven by a common optical field. The mechanical resonators are mechanically coupled through a phonon-hopping interaction via a coupling rate $J_m$  that is modulated through a phase $\theta$. The purpose of such modulation is to allow interference control via the breaking of dark mode in our proposal. Such a dark mode breaking protocol has been recently used to generate quantum correlations in optomechanical systems \cite{Lai_2022,Nori2022}. Based on this interference control, our scheme suggests to further enhance the generated quantum entanglement through quadratic optomechanical coupling. Our  proposed scheme is sketched in \autoref{fig:Fig1}, where both optomechanical and electromechanical analogous setups are displayed. 

In the frame rotating at the driving frequency $\omega_p$, our system is described by the following Hamiltonian ($\hbar=1$),
\begin{equation}\label{eq:ham1}
 H =H_{\rm{0}}+H_{\rm{OM}}+H_{\rm{int}}+H_{\rm{drive}}, 
 \end{equation}
where
\begin{eqnarray}  \label{eq:ham2}
H_{\rm{0}}&:=&\Delta a^\dagger a + \sum_{j=1,2} \omega_m^{j} b_j^\dagger  b_j , \\
H_{\rm{OM}}&:=&- a^\dagger a \sum_{j=1,2} \left[ g_m^{j} (b_j^\dagger + b_j) + g_q^{j} (b_j^\dagger + b_j)^2 \right], \\
H_{\rm{int}}&:=&{J_m}({e^{i\theta}}{b_{1}^{\dagger}}{b_2} + {e^{-i\theta}}{b_1}{b_{2}^{\dagger}}), \\
H_{\rm{drive}}&:=&iE(a^{\dagger} + a).
\end{eqnarray} 
In the above Hamiltonian, $H_{\rm{0}}$, $H_{\rm{OM}}$, $H_{\rm{int}}$ and $H_{\rm{drive}}$ represent respectively the free Hamiltonian, the optomechanical interaction, the mechanical interaction and the driving field. The terms $a$ ($a^\dagger$) and $b_j$ ($b_j^\dagger$) are the annihilation (creation) operators of the electromagnetic mode and for the $j^{th}$ mechanical mode. The frequency of the $j^{th}$ mechanical resonator (cavity) is $\omega_m^j$ ($\omega_c$). The linear and quadratic couplings, between the driving field and the $j^{th}$ mechancial resonator, are captured by  $g_m^j$ and  $g_q^j$, respectively.  The amplitude of the driving field is $E$ and the frequency detuning is defined by $\Delta=\omega_p-\omega_c$. Without loss to the generality, we will assume in the following that our mechancial resonators are degenerated ($\omega_m^j\equiv\omega_m$) and that $g_m^j\equiv g_m$ together with $g_q^j\equiv g_q$ 

The dynamical equations of our system are derived from the Heisenberg equation, which yield,

\begin{equation}\label{eq:qle}
 \begin{cases}
\dot{a}&=\left[i\left(\Delta+ \sum_{j=1,2} \left[g_m(b_j^\dagger + b_j)+g_q(b\dagger + b_j)^2\right]\right)-\frac{\kappa}{2}\right] a  \\&+ E + \sqrt{\kappa}{a}^{in}, \\
\dot{b}_j&=-(i\omega_m +\frac{\gamma_j}{2}){b_j}+ i\left[g_m+2g_q(b_j^\dagger + b_j)\right]a^\dagger a  \\&-iJ_m e^{(-1)^{3-j}i\theta} b_{3-j} + \sqrt{\gamma_j}{b}_j^{in},
 \end{cases}
\end{equation}
where the electromagnetic ($\kappa$) and  mechanical ($\gamma_j$) dissipations have been included. The quantities $a^{in}$ and $b_j^{in}$ stand for the driving and thermal quantum noise operators. These operators are characterized by zero mean values, and the following correlation functions,
\begin{eqnarray}\label{eq:noise}
&\langle a^{in}(t)a^{in\dagger}(t') \rangle &=\delta(t-t'), \hspace*{0.5cm} \langle a^{in\dagger}(t)a^{in}(t') \rangle = 0, \\
&\langle b_j^{in}(t)b_j^{in^\dagger}(t') \rangle &= (n_{th}^j+1)\delta(t-t'),\\ 
&\langle b_j^{in^\dagger}(t)b_j^{in}(t') \rangle &= n_{th}^j\delta(t-t'),
\end{eqnarray}
where $n_{th}^j$ is the thermal phonon occupation number of the $j^{th}$ mechanical resonator that is defined as $n_{th}^j=[\rm exp(\frac{\hbar \omega_j}{k_bT})-1]^{-1}$, with $\rm k_b$ the Boltzmann constant, and $T$ the bath temperature.

To investigate on the quantum entanglement in our system, we need to caputure quantum fluctuations around the mean values of the operators involved in our Quantum Langevin Equations (QLEs) displayed in \autoref{eq:qle}. For this purpose, we linearize our QLEs by using the standard linearization procedure, where the operators are splitted as sum of their mean value and quantum fluctuations. This procedure leads to  the following set of mean value equations,

\begin{equation}\label{eq:means}
\begin{cases}
\dot{\alpha} &=\left(i\tilde{\Delta}-\frac{\kappa}{2}\right)\alpha + E, \\
\dot{\beta}_j &=-(i\omega_m +\frac{\gamma_j}{2}){\beta_j} + i\left[g_m+2g_q(\beta_j^\dagger + \beta_j)\right]|\alpha|^2  \\ &-iJ_m e^{(-1)^{3-j}i\theta} \beta_{3-j} ,
\end{cases}
\end{equation}
together with the set of the fluctuation dynamical equations, 
\begin{equation}\label{eq:fluct}
\begin{cases}
\delta\dot{a}&=\left(i\tilde{\Delta}-\frac{\kappa}{2}\right)\delta a + i\sum_{j=1}^2 \Gamma_j(\delta b_j^\dagger +  \delta b_j) + \sqrt{\kappa}{a}^{in}, \\
\delta\dot{b}_j&=-(i\omega_m +\frac{\gamma_j}{2}){\delta b_j}  + i \Gamma_j\left(\delta a+ \delta a^\dagger\right)\\&+ 2ig_q |\alpha|^2(\delta b_j^\dagger +  \delta b_j) -iJ_m e^{(-1)^{3-j}i\theta} \delta b_{3-j} + \sqrt{\gamma_j}{b}_j^{in},
\end{cases}
\end{equation}
where $\tilde{\Delta}=\Delta+2\sum_{j=1}^2 g_m{\rm Re}(\beta_j)+2g_q({\rm Re}(\beta_j))^2$ is the effective detuning. We have defined $\Gamma_j=G + 2G_q^{j}$, with the following effective linear coupling $G$ ($G=g_m\alpha$) and quadratic coupling  $G_q^j$ ($G_q^j=2g_q\alpha {\rm Re}(\beta_j)$). In the above expressions,   $\alpha$ and $\beta_j$ are the steady-state mean values of the electromagnetic and the $j^{th}$ mechanical modes, respectively. Based on the above fluctuation equations (\autoref{eq:fluct}), we define the following position and momentum quadrature operators, $\delta X_{\mathcal{O}} =\frac{\delta \mathcal{O}^\dagger + \delta \mathcal{O}}{\sqrt{2}}$, $\delta Y_{\mathcal{O}} =i\frac{\delta \mathcal{O}^\dagger - \delta \mathcal{O}}{\sqrt{2}}$, where $\mathcal{O}\equiv a,b_j$, together with their related noise quadratures $\delta X_{\mathcal{O}}^{in} =\frac{\mathcal{O}^{\dagger in} + \mathcal{O}^{in}}{\sqrt{2}}$, $\delta Y_{\mathcal{O}}^{in} =i\frac{\mathcal{O}^{\dagger in} - \mathcal{O}^{in}}{\sqrt{2}}$. This procedure allows us to rewrite the fluctuation dynamical equations in a compact form,
\begin{equation}\label{eq:quadra}
\dot{u}={\rm M} u+z^{in},
\end{equation}
where ${u}=(\delta X_{a},\delta Y_{a}, \delta X_{b_1},\delta Y_{b_1}, \delta X_{b_2},\delta Y_{b_2})^\top$, $z^{in}=(\sqrt{\kappa} X_{a}^{in},\sqrt{\kappa} Y_{a}^{in},\sqrt{\gamma_1} X_{b_1}^{in},\sqrt{\gamma_1} Y_{b_1}^{in},\sqrt{\gamma_2} X_{b_m}^{in},\sqrt{\gamma_2} Y_{b_m}^{in},)^\top$ and the matrix $\rm M$ is given by,
\begin{equation}\label{eq:matrix}
{\rm M}=\begin{pmatrix}
M_1&M_{1|2}&M_{1|3}\\
M_{1|2}^\top&M_2&M_{2|3}\\
M_{1|3}^\top&M_{2|3}^\top&M_3\\
\end{pmatrix},
\end{equation}
with
\begin{eqnarray}
&{\rm M}_1 =\begin{pmatrix}
-\frac{\kappa}{2}&-\tilde{\Delta} \\
\tilde{\Delta}&-\frac{\kappa}{2}
\end{pmatrix}, \hspace*{0.5cm}
{\rm M}_{1+j}=\begin{pmatrix}
-\frac{\gamma_j}{2}&\omega_j \\
-\tilde{\omega}_j&-\frac{\gamma_j}{2}
\end{pmatrix}, \nonumber \\
&{\rm M}_{1|1+j}=\begin{pmatrix}
0&0 \\
2({G}+2G_q^j)&0
\end{pmatrix}, \hspace*{0.1cm}
{\rm M}_{2|3}=\begin{pmatrix}
J_m\sin\theta&J_m\cos\theta \\
-J_m\cos\theta&J_m\sin\theta
\end{pmatrix}. \nonumber 
\end{eqnarray}
In these expressions, the effective frequency $\tilde{\omega}_j=\omega_j-4g_q|\alpha|^2$ for $j=1,2$ has been defined, and the effective couplings $G$ and $G_q^j$ have been assumed to be real for simplicity.
\begin{figure}[tbh]
\begin{center}
  \resizebox{0.5\textwidth}{!}{
  \includegraphics{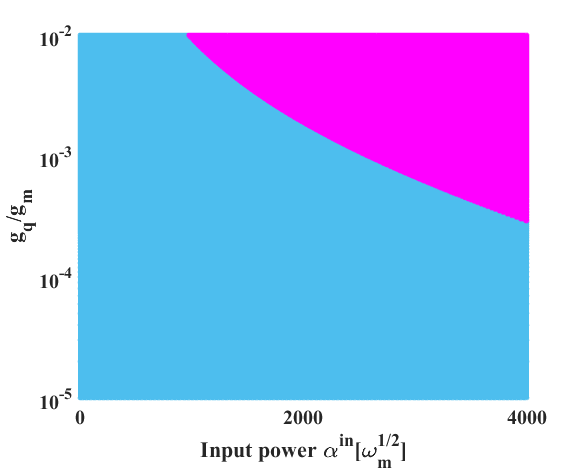}}
  \end{center}
\caption{Stability diagram of the system depending on the input power $\alpha^{in}$ and the optomechanical coupling ratio $g_q/g_m$. The blue/light region is stable, while the pin/dark area represents the unstable zone. The used parameters are $\tilde{\Delta}=-\omega_m$, $g_m=10^{-4}\omega_m$, $\kappa=0.2\omega_m$, $\omega_j=\omega_m$, $\gamma_j=10^{-5}\omega_m$, $J_m=0.1\omega_m$ and $\theta=\pi/2$.}
\label{fig:Fig2}
\end{figure}

\section{Enhancement of optomechanical entanglement}\label{sec:entang}

To analyze the quantum optomechanical entanglement in our system, we need to evaluate the covariance matrix whose elements are defined as $V_{ij}=\frac{\langle u_i u_j  + u_j u_i \rangle}{2}$, which also satisfy the motional equation,
\begin{equation}\label{eq:lyad}
\dot{V}={\rm M}V+V{\rm M^T}+D,
\end{equation}
where $D$ is the diagonal diffusion matrix expressed as $D=\rm{Diag}[\frac{\kappa}{2},  \frac{\kappa}{2}, \frac{\gamma_1}{2}(2n_{th}^{1} + 1), \frac{\gamma_1}{2}(2n_{th}^{1} + 1), \frac{\gamma_2}{2}(2n_{th}^{2} + 1), \frac{\gamma_2}{2}(2n_{th}^{2} + 1)]$. To carry out the entanglement analysis, the matrix $\rm M$ must fulfills the Routh-Huritz stability criterion i.e., all its eigenvalues should have negative real parts \cite{DeJesus}. For this purpose, we have used the following state of the art optomechanical parameters \cite{Fang2016,Fang.2017}, $\omega_m/2\pi=6\rm{GHz}$, $\tilde{\Delta}=-\omega_m$, $g_m=10^{-4}\omega_m$, $\kappa=0.2\omega_m$, $\omega_j=\omega_m$, $\gamma_j=10^{-5}\omega_m$, $J_m=0.1\omega_m$ and $\theta=\pi/2$.  We have checked that these used parameters satisfy the stability condition according to the stability diagram displayed in \autoref{fig:Fig2}. This figure shows how the system is stable (blue/light area) for a wide range of the input field (up to $4\times 10^{3}\omega_m^{1/2}$) and for the quadratic coupling strength up to $g_q/g_m\sim 5\times 10^{-4}$. For $g_q/g_m\gtrsim 5\times 10^{-4}$ however, the unstabilities (pin/dark region) start to strigger into the system as the input power increases.  To capture the steady-state entanglement, it is required that the dynamical variables in \autoref{eq:lyad} do not any more depend on time, and that requirement reduces the above equation to the following Lyaponuv one,

\begin{equation}\label{eq:lyap}
{\rm M}V+V{\rm M^{\top}}=-D.
\end{equation}
The $V_{ij}$ elements of the covariance matrix can be numerically computed, leading to the general form of $V$, 
\begin{equation}\label{cov}
\rm{V}=
\begin{pmatrix} 
V_{\alpha}&V_{\alpha,\beta_1}&V_{\alpha,\beta_2} \\
V_{\alpha,\beta_1}^{\top}&V_{\beta_1}&V_{\beta_1,\beta_2} \\ 
V_{\alpha,\beta_2}^{\top}&V_{\beta_1,\beta_2}^{\top}&V_{\beta_2}
\end{pmatrix},
\end{equation}
where $V_i$ and  $V_{ij}$ are blokcs of $2\times2$ matrices (with $i,j \equiv \alpha,\beta_1,\beta_2$). The diagonal blokcs $V_i$ correspond to the optical mode ($i = \alpha$), and to the mechanical modes ($i = \beta_{1,2}$), respectively. The off-diagonal blocks capture the correlations between the different subsystems. For instance, $V_{\alpha,\beta_1}$ stands for the correlations between the driving field and the first mechanical resonator, $V_{\alpha,\beta_2}$ represents the correlations between the driving field and the second mechanical resonator, while $V_{\beta_1,\beta_2}$ describes the correlations between the mechanical resonators. In our investigation, we will focus on the entanglement between the driving field and the mechanical resonators, since there is no entanglement between the two mechancial resonators (not shown here). To evaluate such bipartite entanglement, we use the logarithmic negativity ($E_N$), which is computed by tracing out the non-necessary third mode. This logarithmic negativity $E_N$ is defined as,
\begin{equation}\label{eq:en}
E_N=\rm max[0,-\ln(2\nu^-)], 
\end{equation}
where $\nu^- = 2^{-1/2}\left[ \Delta_{\chi}-\sqrt{\Delta_{\chi}^2-4I_4}\right]^{1/2}$. The entanglement exists in the system when the condition $\nu^-<1/2$ is satisfied, which is equivalent to Simon's necessary and sufficient criterion for entanglement. The reduced covariance matrix $\chi$, capturing the dynamics of our targeted  optomechanical subsystem is defined as,
\begin{equation}\label{eq:subcov}
\rm{\chi}=\begin{pmatrix} 
V_{\alpha}&V_{\alpha,\beta_j} \\
V_{\alpha,\beta_j}^{\top}&V_{\beta_j}
\end{pmatrix},
\end{equation}
in such a way that $\Delta_{\chi}=\rm{I_1+I_2-2I_3}$, with the following symplectic invariants $I_1=\rm det V_{\alpha}$, $I_2=\rm det V_{\beta_j}$, $I_3=\rm det V_{\alpha,\beta_j}$, and $I_4=\rm det \chi$. In what follows, we assume that our system dwells into the red-sideband regime ($\tilde{\Delta}=-\omega_m$), which is a  requirement for quantum engineering, as heating channels are suppressed in our system.

\begin{figure}[tbh]
\begin{center}
  \resizebox{0.5\textwidth}{!}{
  \includegraphics{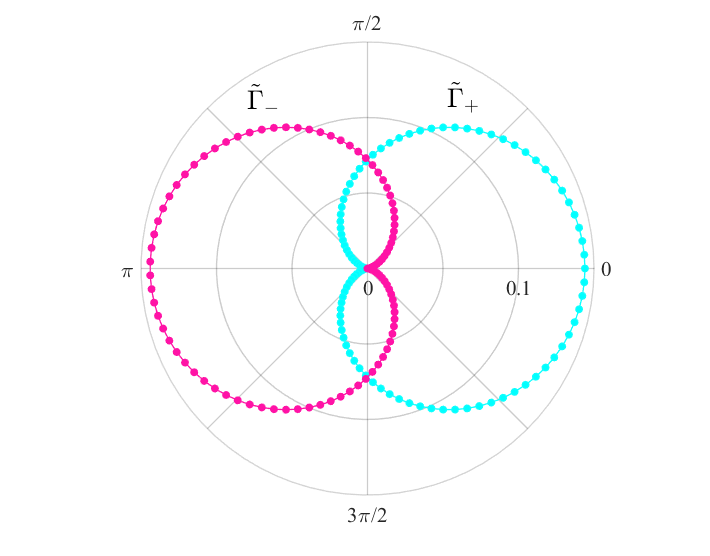}}
  \end{center}
\caption{Redefined coupling strengths $\tilde{\Gamma}_{\pm}$ versus the modulation phase $\theta$ for $\alpha^{in}=2000\omega_m^{1/2}$ and $g_m/g_q=5\times 10^{-4}\omega_m$.  The other parameters are the same as in \autoref{fig:Fig2}.}
\label{fig:Figa}
\end{figure} 

In order to address the dark mode issue in our investigation, we assume that our linearized Hamiltonian under rotating wave approximation (the red-sideband), can be written as  (where the noise terms have been neglected)
\begin{align}
H_{lin}&=-\tilde{\Delta} \delta a^\dagger  \delta a +  \sum_{j=1,2} \left[\tilde\omega_m^j \delta b_j^\dagger \delta b_j  - \Gamma_j(\delta a \delta b_j^\dagger + \delta a^\dagger \delta b_j)\right] \nonumber \\&+{J_m}({e^{i\theta}}{\delta b_{1}^{\dagger}}\delta b_2 + {e^{-i\theta}}{\delta b_1}\delta b_2^{\dagger}),
\end{align}
where we have defined $\tilde \omega_m^j=\omega_m^j-2g_q|\alpha|^2$. 

In the absence of the mechanical coupling ($J_m=0$), bright and dark modes can be caputured  respectively by the modes $B_+$ and a dark $B_-$, which are defined as,
\begin{align}
B_+= \frac{\Gamma_1\delta b_1+\Gamma_2\delta b_2}{\mathcal{G}}, \hspace{1.5em}
B_-= \frac{\Gamma_2\delta b_1-\Gamma_1\delta b_2}{\mathcal{G}}, 
\end{align}
where $\mathcal{G}=\sqrt{\Gamma_1^2+\Gamma_2^{2}}$. The introduction of these terms in the linearized Hamiltonian leads to,
\begin{align} \label{eq:dark0}
H_{lin}&=-\tilde{\Delta} \delta a^\dagger\delta a +\sum_{j=\pm} \tilde\omega_j B_j^\dagger B_j - \Gamma_- (B_+^\dagger B_- + B_-^\dagger B_+)  \nonumber \\&- \Gamma_+(B_+^\dagger \delta a + B_+ \delta a^\dagger)
\end{align}
with $\omega_{+(-)}=\frac{{\tilde\omega}_m^1 \Gamma_{1(2)}^2 + {\tilde\omega}_m^2 \Gamma_{2(1)}^{2}}{\mathcal{G}^2}$, $\Gamma_-=\frac{\Gamma_{1}\Gamma_{2}({\tilde\omega}_m^1-{\tilde\omega}_m^2)}{\mathcal{G}^2}$ and $\Gamma_+=\mathcal{G}$. 
From \autoref{eq:dark0}, it can be seen that under the condition $\tilde \omega_m^1=\tilde \omega_m^2$ (or $\Gamma_-=0$), the mode $B_-$ is decoupled from the system and this contributes to supress some entanglement channels in our proposal. Clearly speaking, this decoupling of mode $B_-$ from the system means that the mechanical dark mode (MDM) has been decoupled from the driving field, resulting in a lack of driven-mechanical entanglement with that mechancial mode. In order to ensure optomechanical entanglement with both mechanical resonators, we introduce the mechanical coupling that brings into play the synthetic magnetism through the phase $\theta$. By taking into account the mechanical coupling ($J_m \neq 0$), the bright and dark modes are captured through,
\begin{align}
\tilde{B}_+&= f\delta b_1-e^{i\theta}h\delta b_2 \\
\tilde{B}_-&= e^{-i\theta}h\delta b_1+f\delta b_2.  
\end{align}
By using these terms in our linearized Hamiltonian, it yields,
\begin{equation}
H_{lin}=-\tilde{\Delta} \delta a^\dagger\delta a +\sum_{j=\pm} \Big( \tilde{\omega}_j \tilde{B}_j^\dagger \tilde{B}_j - \tilde\Gamma_j^{\ast}\delta a \tilde{B}_j^\dagger - \tilde \Gamma_j \delta a^\dagger \tilde{B}_j \Big),
\end{equation}
where $\tilde{\omega}_\pm=\frac{1}{2}\Big(\tilde\omega_m^1 + \tilde\omega_m^2 \pm \sqrt{(\tilde\omega_m^1 - \tilde\omega_m^2)^2 + 4J_m^{2}}\Big)$, and  
\begin{align}\label{eq:Rc}
\tilde{\Gamma}_+=f\Gamma_1 - e^{-i\theta}h \Gamma_2, \hspace{0.5cm}  \tilde{\Gamma}_-=e^{i\theta}h\Gamma_1 + f\Gamma_2,
\end{align}
with $f=\frac{|\tilde{\omega}_- - \tilde\omega_m^1|}{\sqrt{(\tilde{\omega}_- - \tilde\omega_m^1)^2 + J_m^{2}}}$ and $h=\frac{fJ_m}{\tilde{\omega}_- - \tilde\omega_m^1}$. The coupling strengths redefined in \autoref{eq:Rc} are displayed in \autoref{fig:Figa} versus the phase modulation in a polar coordinates. It can be seen that the MDM emerges from the system for $\theta=n\pi$ (for an integer $n$). This situation corresponds to an unbroken dark mode regime, and it can be broken ($\tilde \Gamma_{\pm}\neq0$) by tuning the phase at $\theta \neq n\pi$ for instance, leading to the breaking of the dark mode in the system. Clearly speaking, once that the condition $\theta=n\pi$ is fulfilled, $\tilde B_+$ is a dark mode ($\tilde \Gamma_+=0$) for an odd $n$ (see blue/light curve), while $\tilde B_-$ becomes a dark mode ($\tilde  \Gamma_-=0$) for an even $n$ (see pin/dark curve). Therefore, it appears that coupling the dark mode to the remaining system is only possible through an adjustment of the phase $\theta \neq n\pi$ as it will be seen in the following. The above couplings can be further simplified by assuming $\tilde\omega_m^{j=1,2}=\omega_m$ and $\Gamma_{j=1,2}=\Gamma$ that results to $\tilde{\Gamma}_{\pm}=\frac{\Gamma(1\pm e^{\mp i\theta})}{\sqrt{2}}$, which are strongly $\theta$-dependent as expected. 

\begin{figure}[tbh]
\begin{center}
  \resizebox{0.5\textwidth}{!}{
  \includegraphics{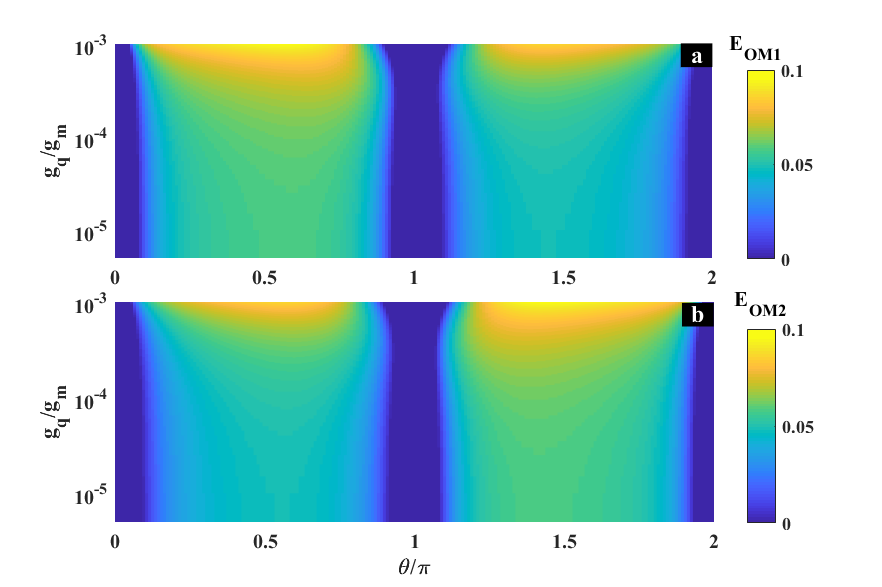}}
  \end{center}
\caption{Contour plot of the optomechanical entanglements versus the phase $\theta$ and the quadratic coupling.  (a) Optomechanical entanglement between the driving field and the first mechanical resonator ($E_{OM1}$). (b) Optomechanical entanglement between the driving field and the second mechanical resonator $E_{OM2}$. We have used $\alpha^{in}=2000\omega_m^{1/2}$ and $n_{th}^j=100$. The others parameters are similar to the ones in \autoref{fig:Fig2}.}
\label{fig:Fig3}
\end{figure}

In what follow, the optomechanical entanglement between the driving field and the first mechanical mode is captured by $E_{OM1}$, while $E_{OM2}$ stands for the optomechanical entanglement between the driving field and the second mechanical mode. \autoref{fig:Fig3} displays the contour plot of the logarithmic negativity versus the phase $\theta$ and the optomechanical coupling ratio $g_q/g_m$. In this figure, \autoref{fig:Fig3}a represents the entanglement between the driving field and the first mechanical resonator, while \autoref{fig:Fig3}b depicts the entanglement between the driving field and the second mechanical resonator.  It can be seen that this figure displays pattern-like interferences with the bright features appearing at $\theta=n\frac{\pi}{2}$ for an odd interger $n$, while the dark features are showing up for $\theta=n\pi$ for an integer $n$. It appears that quantum entanglement is enhanced in our proposal by tunning the phase so that it fulfills the condition $\theta=n\frac{\pi}{2}$ for an odd interger $n$. However, for a phase adjusted at $\theta=n\pi$ for an integer $n$, there is no entanglement in our system. Therefore,  a tuning of the phase $\theta$ can be used to turn on or off the dark/bright mode in our system, providing a path to engineer entanglement through a dark mode breaking process as aforementioned. More importantly, \autoref{fig:Fig3} shows that once the dark mode is broken, the generated entanglement is greatly enhanced as the quadratic coupling increases. This reveals the merit of taking into account the quadratic coupling in our work. Furthermore, this figure reveals that the entanglement exhibits a sort of oscillatory behavior over the phase $\theta$.             
\begin{figure}[tbh]
\begin{center}
  \resizebox{0.5\textwidth}{!}{
  \includegraphics{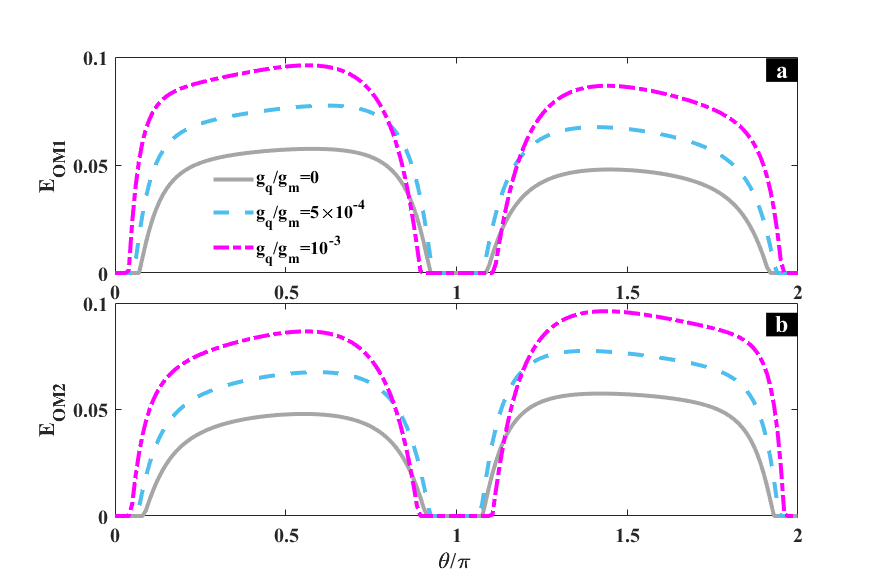}}
  \end{center}
\caption{Optomechanical entanglements versus the phase $\theta$. (a) $E_{OM1}$ versus $\theta$ and (b) $E_{OM2}$ versus $\theta$  for different optomechanical coupling ratio, i.e., $\frac{g_q}{g_m}=0$ (solid line); $\frac{g_q}{g_m}=5\times10^{-4}$ (dashed line); and $\frac{g_q}{g_m}=10^{-3}$ (dash-dotted line). The other parameters are the same as in \autoref{fig:Fig3}.}
\label{fig:Fig4}
\end{figure}

To figure out the oscillatory behavior of the entanglement, we extracted from \autoref{fig:Fig3} the entanglement $E_{OM1}$ and  $E_{OM2}$ versus the phase for different values of the coupling ratio $\frac{g_q}{g_m}$, and they are displayed in \autoref{fig:Fig4}. As aforementioned, it can be seen in \autoref{fig:Fig4} that the entanglement not only exhibits oscillatory behavior, but it is also improved as the quadratic coupling increases.  
       
\begin{figure}[tbh]
  \begin{center}
  \resizebox{0.5\textwidth}{!}{
  \includegraphics{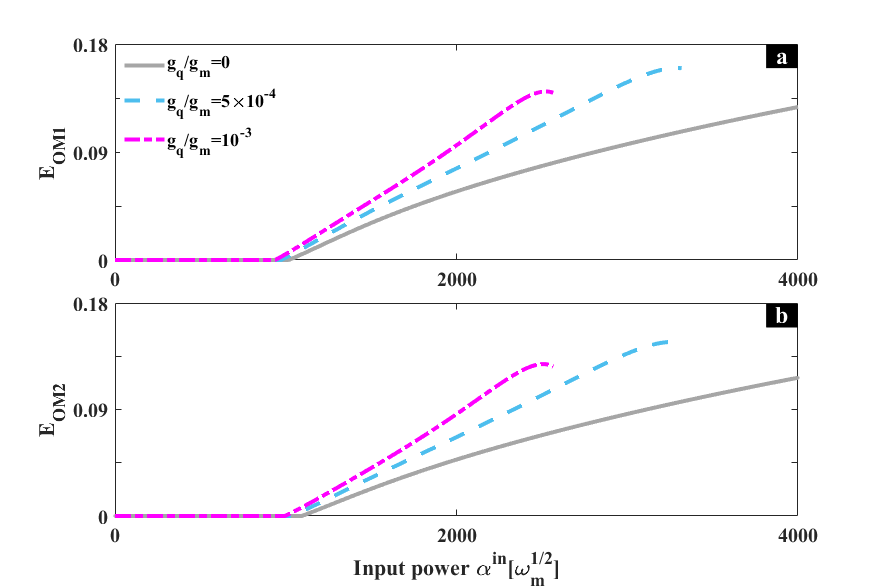}}
  \end{center}
\caption{Optomechanical entanglements, i.e., (a) for $E_{OM1}$ and (b) for $E_{OM2}$, versus the input power $\alpha^{in}$ for different optomechanical coupling ratio, i.e., $\frac{g_q}{g_m}=0$ (solid line); $\frac{g_q}{g_m}=5\times10^{-4}$ (dashed line); and $\frac{g_q}{g_m}=10^{-3}$ (dash-dotted line). The used parameters are the same as those in \autoref{fig:Fig2}.}
\label{fig:Fig5}
\end{figure}

\begin{figure}[tbh]
\begin{center}
  \resizebox{0.5\textwidth}{!}{
  \includegraphics{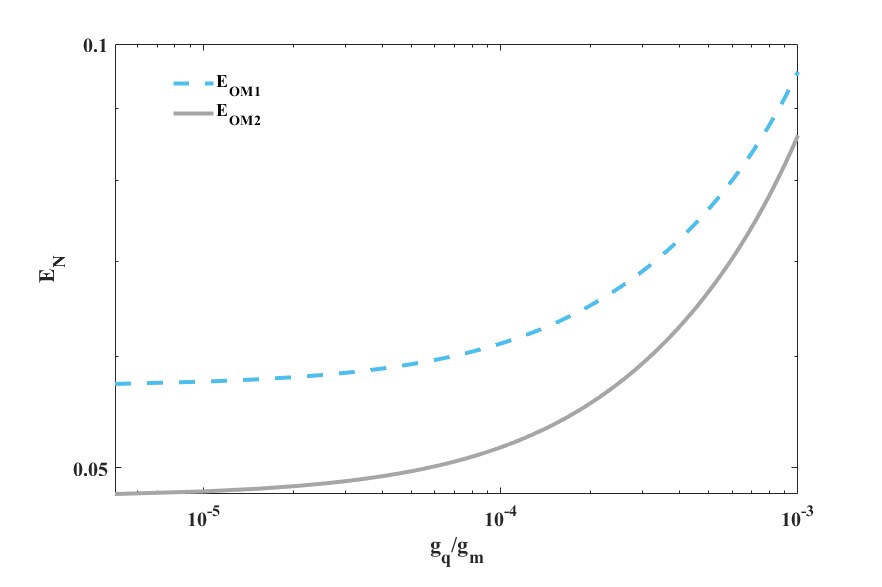}}
  \end{center}
\caption{Optomechanical entanglements, i.e., $E_{OM1}$ (dashed line) and $E_{OM2}$ (solid line), versus the optomechanical coupling ratio $g_q/g_m$ for $n_{th}^j=100$.  The other parameters are the same as in \autoref{fig:Fig2}.}
\label{fig:Fig6}
\end{figure}

In order to see how the entanglement evolves over the input power, and to check our stability analysis, we have plotted  \autoref{fig:Fig5} that captures the dynamics of the optomechanical entanglement versus the input power $\alpha^{in}$ for different coupling ratios $\frac{g_q}{g_m}$. It can be seen that the entanglement exists from a sort of threshold around $\alpha^{in}\sim 10^{3}\omega_m^{-1/2}$. From this threshold, the entanglement strengthens as the input power becomes strong. Moreover, the generated entanglement is further enhanced as the quadratic coupling increases. Furthermore, \autoref{fig:Fig5} shows how the stability starts to matter on the entanglement dynamics as the quadratic coupling is strong enough as it has been displayed in \autoref{fig:Fig2}. Indeed, as shown in \autoref{fig:Fig2}, our system undergoes unstable dynamics when the coupling ratio reaches approximately $\frac{g_q}{g_m}=5\times10^{-4}$, and this agrees well with the unstabilities showing up in \autoref{fig:Fig5}.  In order to further show the dependence of the entanglement $E_{OM1}$ and $E_{OM2}$ with the quadratic coupling, we plotted in \autoref{fig:Fig6} these entanglements versus the coupling ratio $\frac{g_q}{g_m}$. For weak quadratic coupling, \autoref{fig:Fig6} shows that the entanglement is slightly improved. However, when the quadratic coupling is strong enough ($\sim \frac{g_q}{g_m}=10^{-4}$), it can be clearly seen that the entanglement is greatly enhanced.  

\begin{figure}[tbh]
\begin{center}
  \resizebox{0.5\textwidth}{!}{
  \includegraphics{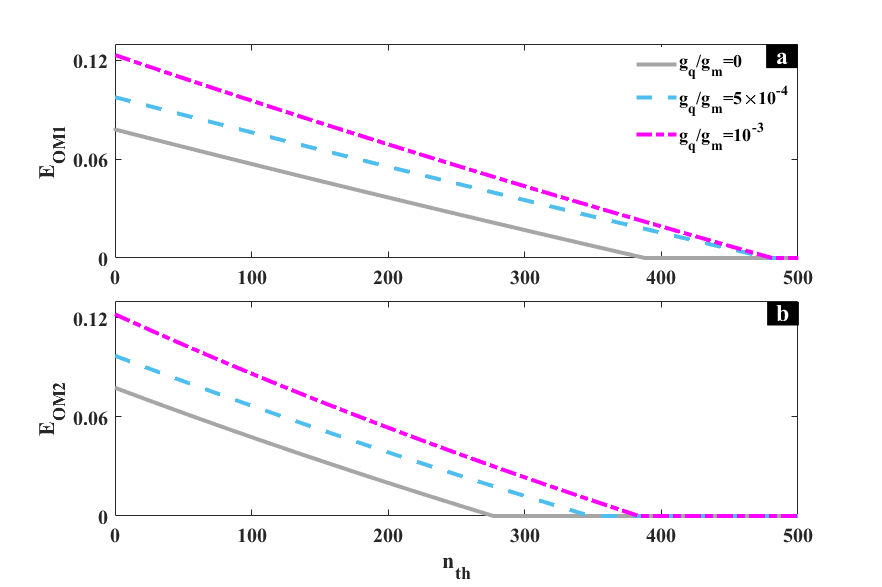}}
  \end{center}
\caption{Optomechanical entanglements, i.e., (a) $E_{OM1}$ and (b) $E_{OM2}$, versus thermal noise $n_{th}$ for different optomechanical coupling ratio, i.e., $\frac{g_q}{g_m}=0$ (solid line); $\frac{g_q}{g_m}=5\times10^{-4}$ (dashed line); and $\frac{g_q}{g_m}=10^{-3}$ (dash-dotted line).  The other parameters are the same as in \autoref{fig:Fig2}.}
\label{fig:Fig7}
\end{figure}

Another interesting feature to figure out is the effect of the thermal noise on the resilience of the entanglement. Such analysis, which reveals the robustness of the entanglement against thermal fluctuations, is displayed in \autoref{fig:Fig7} where it can be seen that the entanglement persists for thermal phonon number up to $n_{th}=400$. This shows that the generated entanglement is robust enough against thermal noise, and this constitutes a good requirement to use these entangled states for quantum information processing, quantum computing, and other quantum tasks.

\section{Conclusion} \label{sec:concl}
This work investigated the effect of the quadratic coupling on the optomechanical entanglement that is induced by a dark mode breaking process. Our benchmark system is made of two mechanically coupled mechanical resonators which are driven by a common electromagnetic field. The mechanical coupling, which is caputured by the phonon hopping rate $J_m$, is modulated through a phase $\theta$ that is used to control the dark mode in our system. By tunning the phase at $\theta=n\frac{\pi}{2}$ with an odd interger $n$, we were able to break the dark mode, leading to an entanglement generation in our proposal. This generated entanglement is greatly enhanced by taking into account the quadratic coupling. Moreover, this entanglement exhibits oscillatory behavior over the phase $\theta$, and the amplitudes of these oscillation like-features increase as the quadratic coupling becomes strong. Furthermore, this generated entanglement is robust enough against thermal noise compared to the case without quadratic coupling. Our scheme can be implemented in electromechanical and optomechanical systems. This generated entangled states can be useful for quantum information processing, and quantum computational tasks.

\section*{Acknowledgments}

This work has been carried out under the Iso-Lomso Fellowship at Stellenbosch Institute for Advanced Study (STIAS), Wallenberg Research Centre at Stellenbosch University, Stellenbosch 7600, South Africa.  K.S. Nisar is grateful to the funding from Prince Sattam bin Abdulaziz University, Saudi Arabia project number (PSAU/2024/R/1445). The authors are thankful to the Deanship of Graduate Studies and Scientific Research at University of Bisha for supporting this work through the Fast-Track Research Support Program.




\bibliography{Squeezing}

\end{document}